\documentclass[10pt,journal,letterpaper,compsoc]{IEEEtran}

\ifCLASSINFOpdf
\else
\fi

\newcommand{\eg}{{\it e.g., }}
\newcommand{\etal}{{\it et~al.}}
\newcommand{\ie}{{\it i.e., }}

\usepackage{balance}
\usepackage{times}
\usepackage{cite} 
\usepackage{pifont}
\usepackage{epsfig}
\usepackage{float}
\usepackage{graphicx} 
\usepackage{amsmath}
\usepackage{url}
\usepackage{subcaption}
\usepackage{amsthm}
\usepackage{algorithmicx}
\usepackage{algpseudocode}
\usepackage{algorithm}
\usepackage{xspace}
\usepackage{comment}
\usepackage{color,soul}

\floatstyle{boxed} 
\newfloat{textfig}{thp}{lop}
\newfloat{textfig}{thp}{lop}
\floatname{textfig}{\textbf{Figure}}
\usepackage{tikz}

\algnewcommand\algorithmicinput{\textbf{Input:}}
\algnewcommand\INPUT{\item[\algorithmicinput]}

\algnewcommand\algorithmicoutput{\textbf{Output:}}
\algnewcommand\OUTPUT{\item[\algorithmicoutput]}

\newlength{\boxfigwidth}

\algdef{SE}[SUBALG]{Indent}{EndIndent}{}{\algorithmicend\ }%
\algtext*{Indent}
\algtext*{EndIndent}

\begin{document}
\title{Performance Analysis and Modeling of Video Transcoding Using Heterogeneous Cloud Services}

\author{Xiangbo Li,
            Mohsen Amini Salehi,~\IEEEmembership{Member,~IEEE,} 
            Yamini Joshi,  \\
            Mahmoud K. Darwich,
            Brad Landreneau, 
            and Magdy Bayoumi,~\IEEEmembership{Fellow,~IEEE} 
\IEEEcompsocitemizethanks{\IEEEcompsocthanksitem Xiangbo Li is with Brightcove Inc. E-mail: xli@brightcove.com
\IEEEcompsocthanksitem Mohsen Amini Salehi, Yamini Joshi, and Brad Landreneau are with the HPCC lab., School of Computing and Informatics, University of Louisiana at Lafayette, LA 70503, USA.\protect\\
 E-mail: \{amini,yxj0845, bml6209\}@louisiana.edu
\IEEEcompsocthanksitem Mahmoud K. Darwich is with the School of Engineering, Math and Technology, Navajo Technical University, NM 87313, USA.\protect\\
 E-mail: mdarwich@navajotech.edu
\IEEEcompsocthanksitem Magdy Bayoumi is with the Department of Electrical and Computer Engineering, University of Louisiana at Lafayette, LA 70503, USA.\protect\\
 E-mail: mab0778@louisiana.edu
    }
 \thanks{}
 }


\IEEEcompsoctitleabstractindextext{

\begin{abstract}

High-quality video streaming, either in form of Video-On-Demand (VOD) or live streaming, usually requires converting (\ie transcoding) video streams to match the characteristics of viewers' devices (\eg in terms of spatial resolution or supported formats). Considering the computational cost of the transcoding operation and the surge in video streaming demands, Streaming Service Providers (SSPs) are becoming reliant on cloud services to guarantee Quality of Service (QoS) of streaming for their viewers. 
Cloud providers offer heterogeneous computational services in form of different types of Virtual Machines (VMs) with diverse prices.
Effective utilization of cloud services for video transcoding requires detailed performance analysis of different video transcoding operations on the heterogeneous cloud VMs. 
In this research, for the first time, we provide a thorough analysis of the performance of the video stream transcoding on heterogeneous cloud VMs. Providing such analysis is crucial for efficient prediction of transcoding time on heterogeneous VMs and for the functionality of any scheduling methods tailored for video transcoding.
Based upon the findings of this analysis and by considering the cost difference of heterogeneous cloud VMs, in this research, we also provide a model to quantify the degree of suitability of each cloud VM type for various transcoding tasks. The provided model can supply resource (VM) provisioning methods with accurate performance and cost trade-offs to efficiently utilize cloud services for video streaming. 
\end{abstract}

}

\maketitle

\IEEEdisplaynotcompsoctitleabstractindextext
\IEEEpeerreviewmaketitle

\section{Introduction}\label{sec:intro}

The way people watch videos has dramatically changed over the past years, from using traditional TV systems to streaming on desktops, laptops, and smartphones through the Internet. Based on the Global Internet Phenomena Report~\cite{intro_1}, video streaming currently constitutes approximately 64\% of all the U.S. Internet traffic. It is estimated that streaming traffic will constitute up to 80\% of the whole Internet traffic by 2019\cite{intro_2}. 

To have a high-quality video streaming experience, video contents, either in the form of Video On Demand (VOD) (\eg YouTube\footnote{https://www.youtube.com} or Netflix\footnote{https://www.netflix.com}) or live-streaming (\eg Livestream\footnote{https://livestreams.com}), need to be converted based on the characteristics of the viewers' devices. That is, the original video has to be converted to a supported resolution, frame rate, video codec, and network bandwidth to match the viewers' display devices~\cite{intro_6}. This conversion is termed \textit{video transcoding}~\cite{intro_7}, which is a computationally-heavy and time-consuming process. 

To minimize the video streaming delay for such diverse viewers, Streaming Service Providers (SSPs) commonly \textit{pre-transcode} videos, \ie they store several versions of the same video~\cite{darwich2016, baik2016vsync}. Given the volume of videos that needs to be transcoded and stored, this approach requires massive storage and processing resources. Provisioning and upgrading built-in infrastructures to meet these demands is cost-prohibitive and distracts SSPs from their mainstream business, which is producing video content and focusing on viewers' satisfaction. Therefore, SSPs have become extensively reliant on cloud providers to provide their services~\cite{pre_3}. The importance and prevalence of video streaming, in addition to its unique QoS demands, has motivated many researchers to investigate dedicated methods for resource allocation and provisioning of video streams (\eg~\cite{li2017cost,rw_10}). 
  
Cloud providers offer abundant of reliable computational and storage services to SSPs. Making use of cloud services, however, poses new challenges to SSPs. One main challenge is to minimize the incurred cost for using cloud services while maintaining QoS (in terms of uninterrupted streaming experience) for their viewers. To overcome this challenge, several research works have been undertaken in estimating video transcoding time~\cite{rw_19, pre_5}, video segmentation models~\cite{bg_2, pre_3}, scheduling~\cite{rw_11, rw_19}, and resource provisioning methods~\cite{pre_3, rw_10}. However, these studies generally focus on elasticity aspect of cloud VMs. That is, how  VMs can be allocated or deallocated to maximize the QoS satisfaction and minimize the incurred cost of SSPs. 

Cloud providers offer a wide variety of VM types (\ie heterogeneous VMs) with diverse prices. For instance, Amazon EC2 offers VM types, such as \texttt{General-Purpose, CPU-Optimized}, and \texttt{GPU} that have different architectural characteristics and remarkably diverse costs. In such a heterogeneous environment, different transcoding operations (also termed \textit{transcoding task}) can potentially have various transcoding times (\ie execution times) on the heterogeneous VMs. The \emph{task-machine affinity} of a task type $i$ on a machine (or VM) type $j$ is defined as how tasks of type $i$ matches (\ie can take advantage of) the architectural characteristics of machine type $j$. Higher affinity implies faster execution time of tasks type $i$ on machine type $j$ \cite{al2011characterizing, maheswaran1999dynamic}.
For instance, particular transcoding tasks can be CPU-intensive whereas some other transcoding tasks can be memory-intensive. More importantly, some transcoding tasks can have similar transcoding times on heterogeneous VMs while their incurred costs vary significantly.

Task scheduling and VM provisioning decisions are critical for SSPs to reduce cost while provide good service. Such decisions should rely on accurate performance information of transcoding tasks and their incurred costs on heterogeneous VMs. Hence, a deep understanding and analysis of the task-machine affinity of transcoding tasks with heterogeneous cloud VMs are required. Currently, there is no study of this kind available yet.

\emph{Expected Time to Compute} (ETC)~\cite{intro_13, intro_14} and \emph{Estimated Computation Speed} (ECS)~\cite{intro_12, intro_15} matrices are commonly used to model and explain the task-machine affinity. However, the definition of both ETC and ECS considers only the execution time as the performance metric and ignores the cost difference across different VM types. The question arises is how we can have a model that captures both the execution time and cost differences of heterogeneous cloud VMs? answering this question can be useful for resource (VM) provisioning methods to allocate appropriate type of VMs for incoming transcoding tasks.

In summary, the \textbf{research questions} we address in this research are:
\textbf{(1)} How can we recognize the task-machine affinity of different transcoding tasks with heterogeneous cloud VMs?
\textbf{(2)} How to model the trade-off between performance and cost of heterogeneous VMs for different transcoding tasks?

To answer the first question, we need to find appropriate factors in video transcoding tasks that can determine the task-machine affinity of transcoding tasks with heterogeneous VMs. In particular, we investigate two factors, namely video transcoding operation and the video content type. 

For that purpose, we analyze the task-machine affinity of transcoding tasks on heterogeneous cloud VMs when the tasks are categorized based on the type of their transcoding operation and when they are categorized based on their content types. However, it is difficult to categorize video transcoding tasks based on their content type because the content type is not known prior to the execution of the tasks. Hence, in the next step, we find factors that indicate the video content type, as such, can be used for categorizing video transcoding tasks. 

To answer the second research question, we present a model to quantify the suitability of heterogeneous VMs for a given transcoding task. The model encompasses both the execution time of the task on a VM type and the incurred cost of using it. 

In summary, the key \textbf{contributions} of this paper are:
\begin{itemize}
\setlength\itemsep{1em}
 \item Analyzing the performance of different transcoding operations on heterogeneous cloud VMs.
 \item Analyzing the performance of video content types on the on heterogeneous cloud VMs.
 \item Determining influential factors on the execution time of the transcoding operation.
 \item Providing a model to capture (and quantify) the cost and performance trade-off of heterogeneous VMs for video transcoding tasks. 
\end{itemize}

The rest of the paper is organized as follows. Section~\ref{sec:bg} provides background on video stream structure, video transcoding and the heterogeneous cloud VMs. In Section~\ref{sec:pe}, we compare and analyze the task-machine affinity of transcoding tasks on heterogeneous cloud VMs. Suitability of transcoding tasks for heterogeneous VMs is proposed in Section~\ref{sec:md}. Section~\ref{sec:rw} reviews related works in the literature and position our work with respect to them. Finally, Section~\ref{sec:conclusion} summarizes findings of the paper.

\section{Background}\label{sec:bg}
\subsection{Video Stream Structure}\label{vbs}
\begin{figure}[htb] 
    \centering
    \includegraphics[width=3.5in]{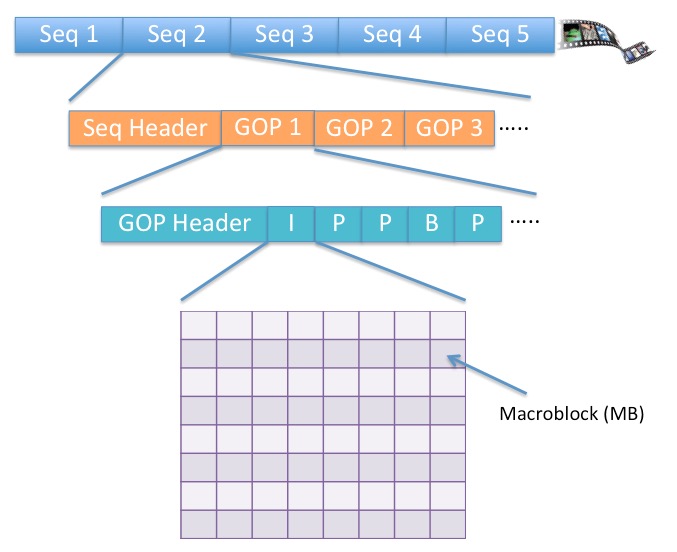}
    \caption{The structure of a video stream. Each sequence includes multiple GOPs. Frames of a GOP are of \texttt{I} (intra), \texttt{P} (predicted), or \texttt{B} (be-directional predicted) types.}
    \label{video_segment}
\end{figure}

A Video stream, as shown in Figure~\ref{video_segment}, consists of several sequences. Each sequence is divided into multiple \textit{Group Of Pictures} (GOP) with sequence header information in the beginning of each GOP. A GOP is essentially a sequence of frames related to the same scene in the video. A GOP starts with an \texttt{I} (intra) frame, followed by a number of \texttt{P} (predicted) frames or \texttt{B} (be-directional predicted) frames~\cite{wiegand2003overview}. 
Each frame contains several \textit{slices} that consist of a number of \textit{macroblocks} which is the unit for video encoding and decoding operations. 
As each GOP can be processed independently, transcoding operation is commonly carried out at the GOP level~\cite{bg_14}. Similarly, in this work, we assume that all the transcoding processes operate at the GOP level.

\subsection{Video Transcoding}\label{sec:vt}
Video contents are initially captured with a particular format, spatial resolution, frame rate, and bit rate. SSPs usually have to adjust the original video based on the viewer's network bandwidth, device resolution, frame rate, and video compression standard (\ie codec). These conversions are carried out on all GOPs of a video and are termed \textit{video transcoding}~\cite{intro_6, intro_7}. Transcoding process includes decoding GOPS and re-encoding them in the new format. Accordingly, transcoding time is the sum of decoding and re-encoding times~\cite{seo2010load}.

Below, we provide more details on the nature of processing in different transcoding operations:

\subsubsection{Bit Rate Adjustment}
To stream high-quality video contents, the videos are encoded with a high bit rate. However, high bit rate also means the video content needs a larger network bandwidth for transmission. Considering the diversity and fluctuations of network bandwidth on the viewer's side, SSPs usually need to change the bit rate of video streams to ensure smooth streaming~\cite{bg_3}. Dynamic bit rate adjustment of video streams is also known as Adaptive video streaming~\cite{jiang2012improving}.

\subsubsection{Spatial Resolution Reduction}
The spatial resolution indicates the dimensional size of a video. The dimensional size of an original video stream does not necessarily match to the screen size of viewers' devices. Thus, to avoid losing content, macroblocks of an original video have to be removed or combined (\ie downscaled) to produce lower spatial resolution video. There are also circumstances where the spatial resolution algorithms can be applied to reduce the spatial resolution without sacrificing quality~\cite{bg_7}.  
 
\subsubsection{Temporal Resolution Reduction}
Temporal resolution reduction happens when the viewer's device only supports a lower frame rate. In this situation, the SSP has to drop some frames. Due to the dependency between frames, dropping frames may cause motion vectors become invalid for the incoming frames. Details of methods for temporal resolution reduction can be found in~\cite{bg_12}.

\subsubsection{Video Compression Standard Conversion}
There is a wide variety of video compression standards (codec) for video files ---from MPEG2~\cite{haskell1996digital}, to H.264~\cite{wiegand2003overview}, and to the most recent one, HEVC~\cite{sullivan2012overview}. Without these compression standards in place, the video size would be too large and cannot be streamed or even stored using the current network and storage capacities. 
Viewer's devices usually support only one or few compression standards. Hence, if the video codec is not supported on the viewer's device, then the video needs to be transcoded based on the supported codec on  the viewer's device~\cite{bg_11}.

\subsection{Video Content Type}
Each GOP covers one scene in a video and utilizes still background content in the video to reduce its size.
Accordingly, based on the frequency of scene changes, video contents can be categorized into three types: \emph{slow motion}, \emph{fast motion}, and \emph{mixed motion}. 

In \emph{slow motion} videos, the scene changes slowly and the background remains still. Therefore, GOPs of such videos include many frames and are large in size.
In contrast, the scene changes of \emph{fast motion} videos (\eg action movies) are dramatic. These videos contain many GOPs, however, each GOP includes few frames, hence, it is small in size.
A \emph{mixed motion} video includes a combination of both fast and slow motion scenes, thus, includes GOPs with a variety of sizes.

\subsection{Heterogeneous VMs in Cloud}\label{subsec:pricing}
Cloud service providers offer heterogeneous computational services (VMs) to satisfy various types and levels of computational requirements of their clients. Heterogeneity of these VMs is based on both underlying hardware characteristics and their hourly cost. Such heterogeneity enables cloud users to build a cluster of heterogeneous VMs to  process high performance computations in the cloud. 
Heterogeneous systems are categorized as \emph{consistent} and \emph{inconsistent}~\cite{intro_12} environments. The former refers to environments in which some machines (VMs) are faster than others whereas the latter explains an environment in which tasks have diverse execution times on heterogeneous
machines. For instance, machine A may be faster than machine B for task 1 but slower than other machines for task 2~\cite{rw_17}. We also say that machine A has a higher affinity with task 1. 
In fact, cloud providers offer several categories of VMs that are inconsistently heterogeneous. Nevertheless, there is a consistent heterogeneity within VMs in each one of those categories. In this study, our goal is to study the affinity of different transcoding tasks on heterogeneous VMs, thus, we consider a cloud as  inconsistently heterogeneous environments.

In the case of Amazon EC2 cloud, six categories of VM types are offered that are described below: 
\begin{itemize}
\item \textbf{General Purpose VMs:}
This VM type has a fair amount of CPU, memory, and networks for many applications, such as web servers and small- or mid-size database servers. General-purpose VMs are the least expensive one and have lower computing power in comparison with other VM types. Generally, to process a large set of tasks, either many or few of these VMs should be allocated for a long time~\cite{bg_15}. 

\item \textbf{CPU Optimized VMs:}
This VM type offers a higher processing power in comparison with other VM types, which makes them ideal for compute-intensive tasks. They are currently mostly applied for high-traffic web application servers, batch processing, video encoding, and high performance computing applications (\eg genome analysis and high-energy physics)~\cite{bg_16}. 

\item \textbf{Memory Optimized VMs:}
Memory-Optimized VM type is designed for processing tasks with large memory demand. This VM type has the lowest cost per GB of memory (RAM) compared to other types. Applications such as high performance databases, distributed cache, and memory analytics~\cite{bg_18} usually demand Memory-Optimized VMs.

\item \textbf{GPU Optimized VMs:}
The GPU-Optimized VMs are applied for compute-intensive tasks (\ie tasks that involve huge mathematical operations). Many large-scale simulations, such as computational chemistry, rendering, and financial analysis are conducted on GPU-Optimized VMs~\cite{bg_19}.
 
\item \textbf{Storage Optimized and Dense Storage VMs:}
These VM types are utilized in cases where low storage cost and high data density is necessary. This VM type is designed for large (big) data requirements such as Hadoop­ clusters and data warehousing applications~\cite{bg_17}.
 
\end{itemize}

We perform this research by using VM types offered by the Amazon cloud provider. The reason we chose Amazon is that it is the mainstream cloud provider and many video SSPs utilize its services~\cite{adhikari2012unreeling}. However, we would like to note that the analysis provided in this work is general and can be applied to any heterogeneous computing (HC) environment.

\section{Performance Analysis of Transcoding Operations on Heterogeneous Cloud VMs}\label{sec:pe}

\subsection{Overview}
To keep the generality and to avoid limiting the research to the details of VM types offered by Amazon EC2, we select one VM type from different VM categories in Amazon EC2 that represents the characteristics of that category (see Section~\ref{subsec:pricing}). 

In particular, for the General-Purpose, CPU-Optimized, Memory-Optimized, and  GPU VM types we choose \texttt{m4.large}, \texttt{c4.xlarge}, \texttt{r3.xlarge}, and \texttt{g2.2xlarge}, respectively. We did not consider any of the Storage-Optimized and Dense-Storage VM types in our evaluations as we observed that IO and storage are not influential factors for video transcoding tasks. The characteristics and the cost of the chosen VM types are illustrated in Table~\ref{tbl:cost-table}. In this table, vCPU represents virtual CPU. Amazon uses what it calls ``EC2 Compute Units'' or ECUs, as a measure of virtual CPU power. It defines one ECU as the equivalent of a 2007 Intel Xeon or AMD Opteron CPU running at 1 GHz to 1.2 GHz clock rate. More details about the characteristics of the VM types can be found at Amazon EC2 website\footnote{https://aws.amazon.com/ec2/instance-types/}. 

\renewcommand{\arraystretch}{1.5}
\begin{table}[htb]
  \caption{\label{tbl:cost-table}Cost of heterogeneous VMs in Amazon EC2 cloud.}
  \centering{
    \resizebox{\columnwidth}{!}{
      \begin{tabular}{ c | c | c | c | c }
       VM Type & \shortstack{General \\ (\texttt{m4.large})} & \shortstack{CPU Opt.\\ (\texttt{c4.xlarge})} & \shortstack{Mem. Opt.\\ (\texttt{r3.xlarge})} & \shortstack{GPU\\ (\texttt{g2.xlarge})}\\ \hline
       vCPU & 2 & 4 & 4 & 8 \\
       Memory (GB) & 8 & 7.5 & 30.5 & 15 \\
       Hourly Cost (\$) & 0.15 & 0.20 & 0.33 & 0.65 \\ 
      \end{tabular}
    }
  }
\end{table}

To analyze the transcoding time, we utilized a set of benchmark videos. The benchmarking videos are publicly available for reproducibility purposes\footnote{The videos can be downloaded from: https://goo.gl/TE5iJ5}. Videos in the benchmark are diverse both in terms of the content types and length. The benchmark includes a combination of slow, fast, and mixed motion video content types. The length of the videos in the benchmark varies in the range of [10, 600] Seconds. The size and frame number of the benchmark videos ranges from 5MB to 313MB, and 240 to 10464, respectively.

We used \texttt{FFmpeg}\footnote{https://ffmpeg.org}, which is an open source utility, to transcode the videos. State-of-the-art FFmpeg transcoder is a cascaded transcoder with sequential transcoding algorithm, that means the incoming source video stream is fully decoded before re-encoding into the target video stream with the desired codec, bitrate, and frame rate~\cite{intro_6,intro_7}. For each one of the benchmarking videos, four different transcoding operations, namely codec conversion, resolution reduction, bit rate adjustment, and frame rate reduction were carried out on heterogeneous VMs. 

Each transcoding operation has been repeated for 30 times on each video to remove any randomness (\eg due to VM malfunctioning or other temporal issues)\footnote{The workload traces are available at: https://goo.gl/B6T5aj}. The mean transcoding time on each VM for a given GOP is considered for comparison and analysis of this paper.

\subsection{Analyzing the Execution Time of Different Video Transcoding Operations}\label{sub:tt}

\begin{figure*}[htbp]
    \centering
    \begin{subfigure}[b]{0.245\textwidth}
      \centering
      \includegraphics[width=\linewidth]{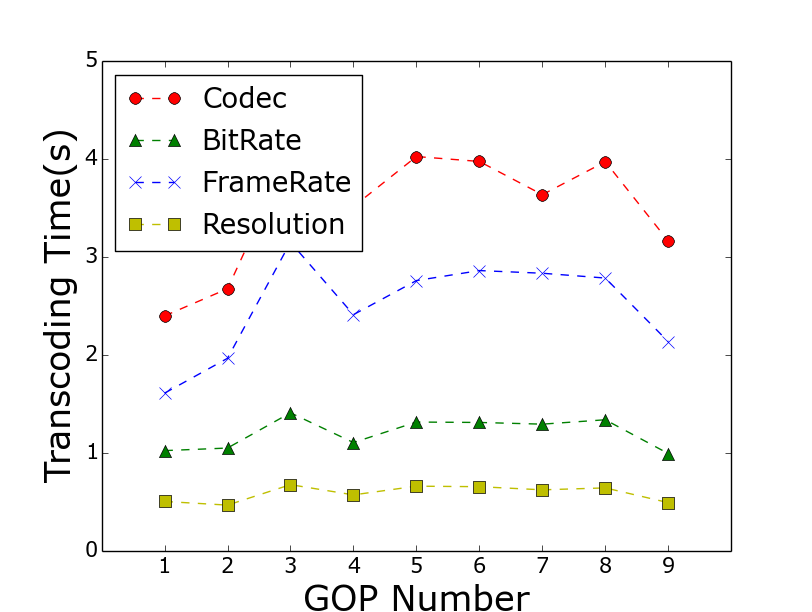}
      \caption{\texttt{General} VMs}
      \label{fig:tm4}
    \end{subfigure}
    \begin{subfigure}[b]{0.245\textwidth}
      \centering
      \includegraphics[width=\linewidth]{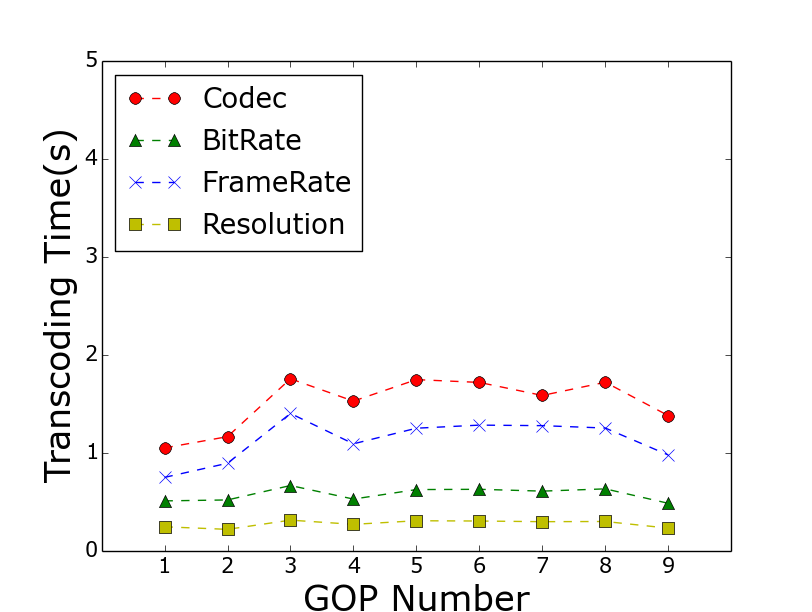}
      \caption{\texttt{CPU Opt.} VMs}
      \label{fig:tc4}
    \end{subfigure}
    \begin{subfigure}[b]{0.245\textwidth}
      \centering
      \includegraphics[width=\linewidth]{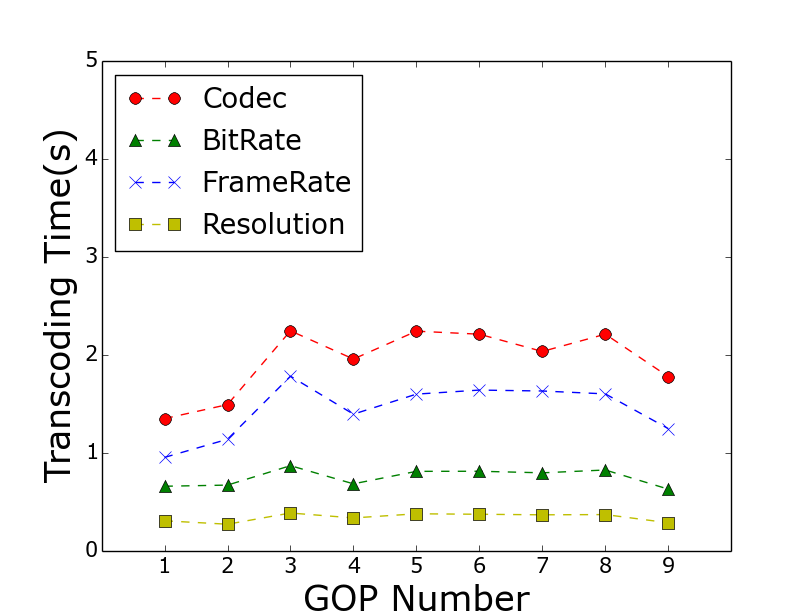}
      \caption{\texttt{Mem. Opt.} VMs}
      \label{fig:tr3}
    \end{subfigure}
    \begin{subfigure}[b]{0.245\textwidth}
      \centering
      \includegraphics[width=\linewidth]{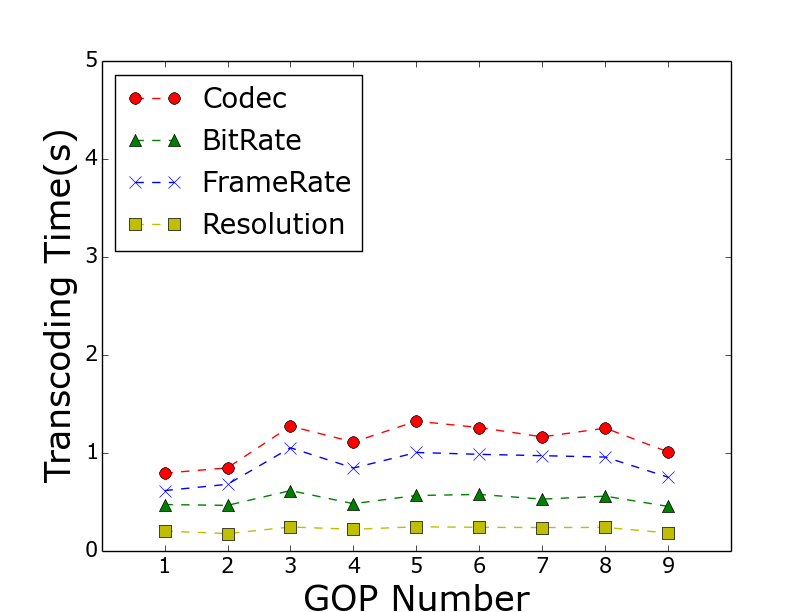}
      \caption{\texttt{GPU} VMs}
      \label{fig:tg2}
    \end{subfigure}
    \caption{Mean transcoding time (in Seconds) on GOPs of benchmark videos on distinct VM types. (a) mean transcoding time of different transcoding operations on \texttt{General} VM. (b), (c), and (d) show mean transcoding time of different transcoding operations using \texttt{CPU Opt., Mem. Opt.,} and \texttt{GPU}, respectively.}
    \label{fig:ts}
\end{figure*}

The first question we need to answer is to identify if a certain transcoding operation has a stronger task-machine affinity with a particular cloud VM type.

To answer this question, we compared the transcoding time (execution time) of various transcoding operations using different VM types. We measured the transcoding time of the first nine GOPs in all videos in the benchmark on different VM types and reported and the mean of their transcoding times. The reason we choose nine GOPs is that the shortest video exists in the benchmark has nine GOPs. We should note that, because GOPs are transcoded independently and there are diverse types of video contents in the repository, the nine GOPs are representative of other GOPs in the benchmark.

Figure~\ref{fig:ts} shows the transcoding time of different transcoding operations on heterogeneous VMs. We can observe that the execution times of different transcoding operations are not the same, however, regardless of the VM type, they follow the same pattern. Sub-figures~\ref{fig:tm4},~\ref{fig:tc4},~\ref{fig:tr3}, and~\ref{fig:tg2} demonstrate that although the execution time of each transcoding operation varies on different VM instances, in general, transcoding time has the same pattern across \texttt{General}, \texttt{CPU Opt.}, \texttt{Mem. Opt.} and \texttt{GPU} VM types. 

The results confirm that, regardless of the VM type utilized, converting video codec always takes more time than other transcoding operations. This is because changing codec implies decoding the original format of the video and then, encoding it to a new codec. These conversions make the transcoding time longer. We also observe that changing resolution has the least transcoding time regardless of the VM type. The reason is that the transcoding is achieved by utilizing filtering and subsampling~\cite{bg_4,yin2000video} which works directly in the compressed domain and avoids the computationally expensive steps of converting to the pixel domain. Therefore, it takes less time than other transcoding operations.

\subsection{Analyzing the Task-Machine Affinity of Video Transcoding Operations with Heterogeneous VMs}\label{sub:it}
\begin{figure}[htb] 
    \centering
    \includegraphics[width=3.5in]{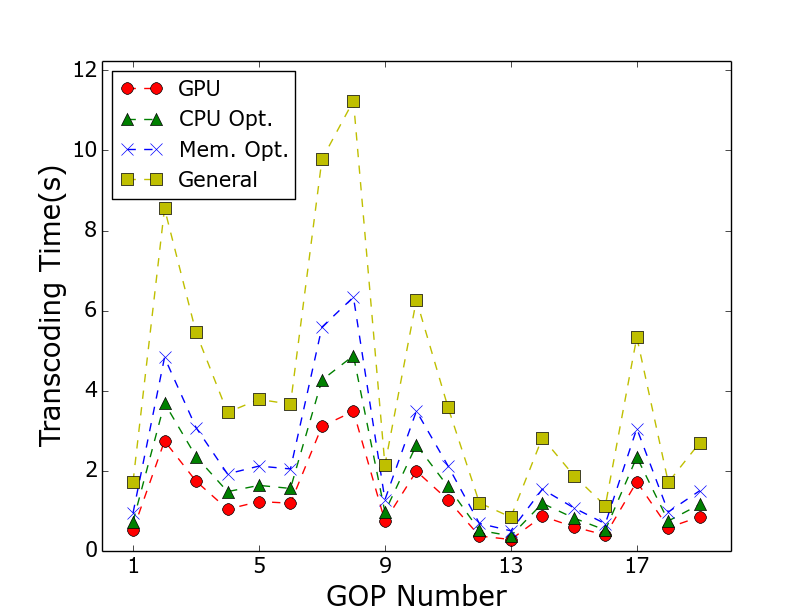}
    \caption{Execution time of codec transcoding for a video in the benchmark on different VM types.}
    \label{dif_instance}
\end{figure}

As mentioned in Section~\ref{subsec:pricing}, cloud providers offer VMs that are heterogeneous both in terms of performance and cost. 
An important question for video stream providers to reduce their cost and improve their Quality of Service (QoS) is: what is the task-machine affinity of video transcoding operations with heterogeneous cloud VMs? 

As we noticed in Section~\ref{sub:tt}, although execution times of various video transcoding operations are different, codec transcoding has the highest execution time and changing spatial resolution generally has the lowest execution time. Considering this pattern, to study the task-machine affinity of video transcoding on heterogeneous VMs, we only consider one transcoding operation (\eg codec transcoding) on heterogeneous VMs. Hence, we measure the codec transcoding time of benchmark videos on heterogeneous VM types. 

Figure~\ref{dif_instance} expresses the analysis for one video\footnote{This is \texttt{big\_buck\_bunny\_720p\_h264\_02tolibx264} video in the benchmark.} in the benchmark. Graph of the same evaluation is illustrated in APPENDIX A for other videos of the benchmark. In this figure, we can observe that, in general, \texttt{GPU} VM provides a better execution time in comparison with other VM types. This is because transcoding operations include substantial mathematical operations and \texttt{GPU} VM types are well suited for such kind of operations. \texttt{General} VM provides the lowest performance as it includes less powerful processing units (see Table~\ref{tbl:cost-table}).

\begin{figure*}[htbp]
    \centering
    \begin{subfigure}[b]{0.32\textwidth}
      \centering
      \includegraphics[width=\textwidth]{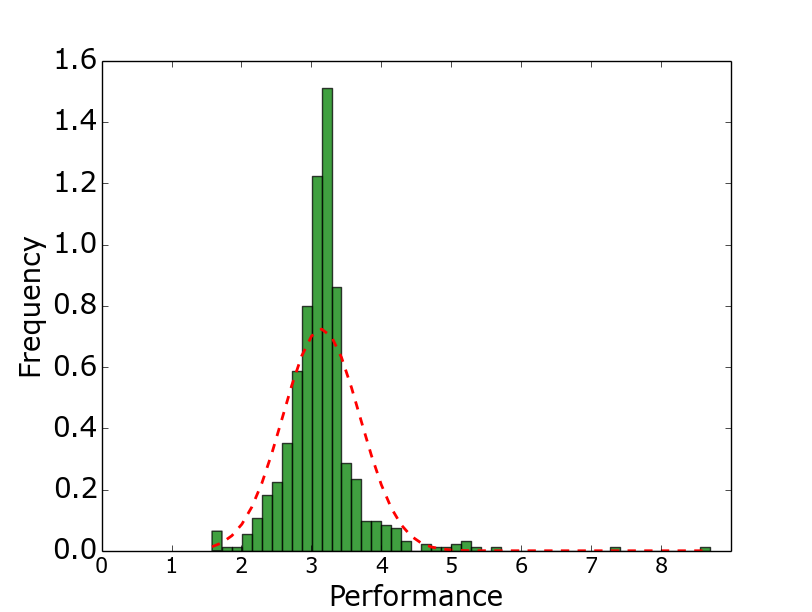}
      \caption{Performance ratio of \texttt{General} to \texttt{GPU}.}
      \label{fig:m4vsg2}
    \end{subfigure}
    \begin{subfigure}[b]{0.32\textwidth}
      \centering
      \includegraphics[width=\textwidth]{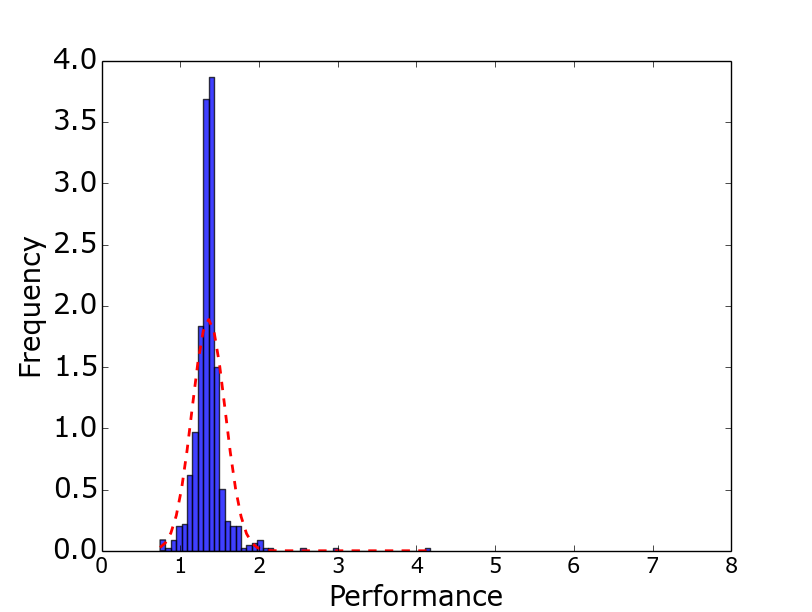}
      \caption{Performance ratio of \texttt{CPU-Opt.} to \texttt{GPU}.}
      \label{fig:c4vsg2}
    \end{subfigure}
    \begin{subfigure}[b]{0.32\textwidth}
      \centering
      \includegraphics[width=\textwidth]{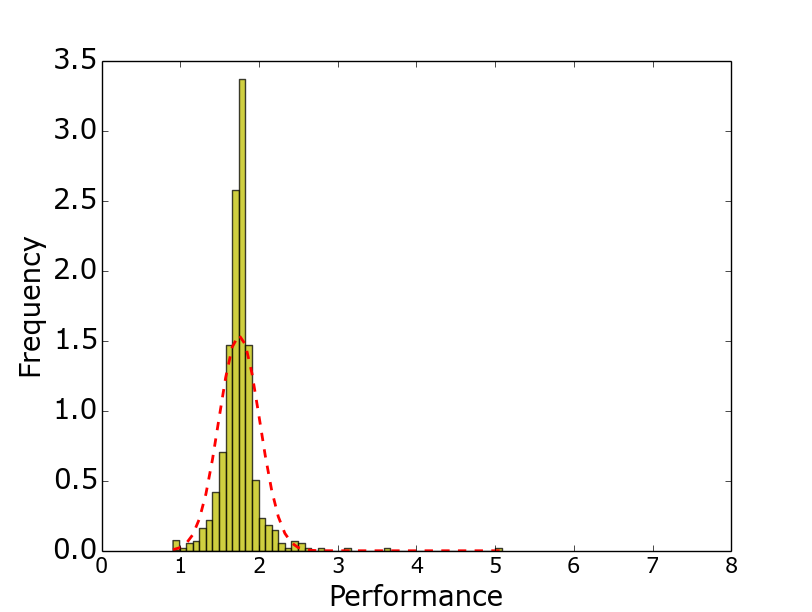}
      \caption{Performance ratio of \texttt{Mem-Opt.} to \texttt{GPU}.}
      \label{fig:r3vsg2}
    \end{subfigure}
    \caption{Performance ratio of transcoding on different VM types with respect to \texttt{GPU} VM. Horizontal axis shows the performance ratio and the vertical axis shows the frequency of the performance ratio for all GOPs of videos in the benchmark.}
    \label{fig:vmtype}
\end{figure*}

More importantly, in Figure~\ref{dif_instance}, we observe that the transcoding times of different GOPs significantly varying on the four VM types. For some GOPs, the \texttt{GPU} VM remarkably outperforms other VMs (\eg GOP 6, 7, and 8) whereas for some other GOPs (\eg GOP 9, 12, and 13) the difference in transcoding times is negligible. 

To better understand the performance variations in transcoding different GOPs, we compared the performance of these four VM types for all videos in the benchmark in detail. Although \texttt{GPU} takes the least time to perform a transcoding operation, we are interested to know the significance of the outperformance of the \texttt{GPU} across different GOPs. Thus, we normalized the transcoding time of GOP $i$ on a given VM type, by dividing it by the transcoding time of GOP $i$ on \texttt{GPU}. The result of this analysis is shown in Figure~\ref{fig:vmtype}.
In all sub-figures of Figure~\ref{fig:vmtype}, the horizontal axis shows the performance ratio and the vertical access shows the frequency of that ratio across all GOPs in a video. That is, the number of times each performance ratio has occurred for all GOPs. We fit a Bell curve on the histograms of these sub-figures and the results conform with the Normal distribution. Mean and Standard Deviation of the fitted Normal distribution, are as follows:

\begin{enumerate}
\item In Sub-figure~\ref{fig:m4vsg2}, performance ratio of \texttt{General} VM lies within the range $ 2.781 \pm 1.524 $.
\item In Sub-figure~\ref{fig:c4vsg2}, performance ratio of \texttt{CPU Opt.} VM lies within the range $ 1.263 \pm 0.508 $. 
\item In Sub-figure~\ref{fig:r3vsg2}, performance ratio of \texttt{Mem. Opt.} VM lies within the range $ 1.608 \pm 0.652 $. 
\end{enumerate}

\renewcommand{\arraystretch}{1.5}
\begin{table}[htb]
  \caption{\label{tbl:perf-table-1}Performance ratio of different VM types with respect to \texttt{GPU} VM, for all GOPs of the videos in the benchmark. Each entry shows the percentage of GOPs with performance ratio $< 1.0$.}
  \centering{
    \resizebox{\columnwidth}{!}{
      \begin{tabular}{ c | c | c | c | c }
        
         & Codec & Frame Rate & Bit Rate & Resolution \\ \hline
        General & 0\% & 2.4\% & 2.7\% & 2.4\% \\ 
        CPU Opt. & 2.2 \% & {\bf24.8}\% & {\bf28.0}\% & 3.9\% \\ 
        Mem. Opt. & 0.6\% & 2.8\% & 4.2\% & 2.5\% \\ 
        
      \end{tabular}
    }
  }
\end{table}

\renewcommand{\arraystretch}{1.5}
\begin{table}[htb]
  \caption{\label{tbl:perf-table-2}Performance ratio of different VM types with respect to \texttt{GPU} VM, for all GOPs of the videos in the benchmark. Each entry shows the percentage of GOPs with performance ratio $\le 1.2$.}
  \centering{
    \resizebox{\columnwidth}{!}{
      \begin{tabular}{ c | c | c | c | c }
        
         & Codec & Frame Rate & Bit Rate & Resolution \\ \hline
        General & 0\% & 2.72\% & 2.87\%  & 2.72\% \\
        CPU Opt. & 12.28\% & {\bf33.33}\% & {\bf63.93}\% & 22.28\% \\
        Mem. Opt. & 1.36\% & 23.78\% & 23.63\% & 3.49\% \\
        
      \end{tabular}
    }
  }
\end{table}

We also measured the percentage of GOPs transcoded on VMs other than \texttt{GPU} with performance ratio $< 1.0$. That is, the percentage of GOPs that their transcoding time is less than the transcoding time on the \texttt{GPU}. The results are shown in Table~\ref{tbl:perf-table-1}. We see that the percentage of GOPs that have transcoding time strictly lower than the \texttt{GPU} for different transcoding operations. In addition, to see the percentage of transcoding tasks that have close execution time to the \texttt{GPU}, in Table~\ref{tbl:perf-table-2}, the percentage of tasks that have performance ratio lower than $1.2$ are reported. 

\noindent{\textbf{Summary of our observations in this part}}:
\begin{enumerate}
\item We observe that in cases that transcoding time of other VM types are lower than \texttt{GPU}, the transcoding time differences are low (less than 0.24 seconds). We note that 0.24 second is relatively low  and negligible when compared with the delay caused by network.

\item In all cases that other VM types outperform the \texttt{GPU} VM, the transcoding time on the \texttt{GPU} was low (less than 2.1 seconds). That is, when the \texttt{GPU} takes a low time to transcode a GOP, other VM types may outperform it. 

\item From the two previous observations, we conclude that, in a cloud environment with heterogeneous VMs, making use of expensive VM types for tasks with short execution time is not beneficiary. However, understanding the exact execution time threshold requires benchmarking in that particular context and study the performance cost ratio of using different VM types.

\item According to Figure~\ref{fig:vmtype}, none of the transcoding types need extensive memory space (\ie transcoding is not a memory intensive operation). Therefore, video stream providers would not benefit from instantiating memory-optimized VM types for video transcoding.
\end{enumerate}

\subsection{Analyzing the Impact of Video Content Type on Transcoding Time}\label{sub:vt}

\begin{figure*}[htbp]
    \centering
    \begin{subfigure}[b]{0.32\textwidth}
      \centering
      \includegraphics[width=\textwidth]{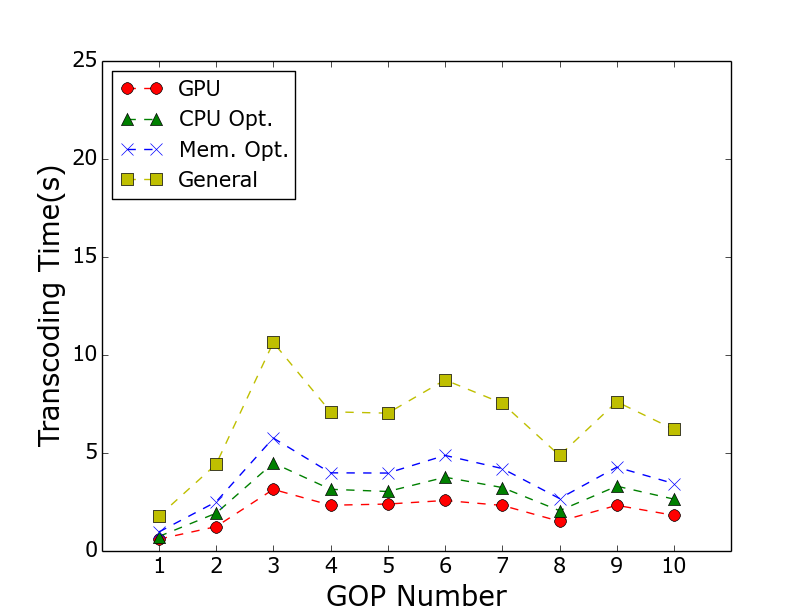}
      \caption{Slow motion video.}
      \label{fig:slow}
    \end{subfigure}
    \begin{subfigure}[b]{0.32\textwidth}
      \centering
      \includegraphics[width=\textwidth]{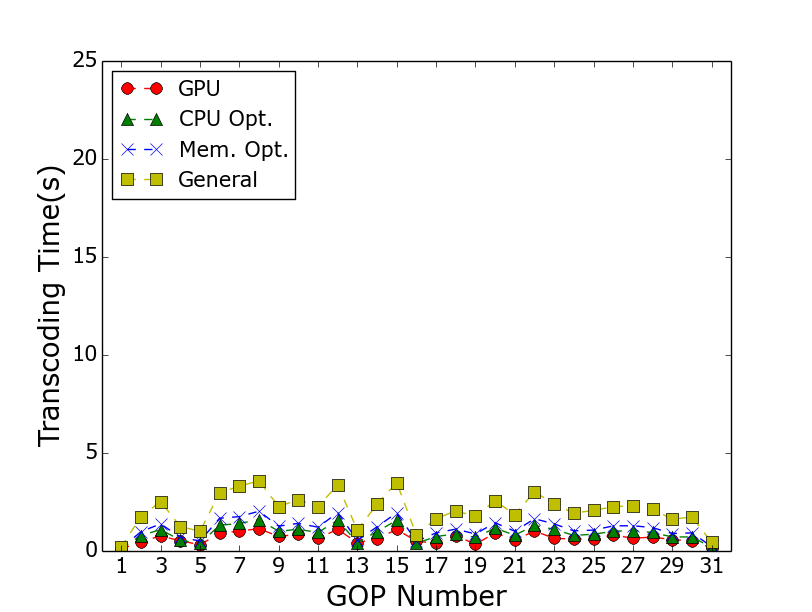}
      \caption{Fast motion video.}
      \label{fig:fast}
    \end{subfigure}
    \begin{subfigure}[b]{0.32\textwidth}
      \centering
      \includegraphics[width=\textwidth]{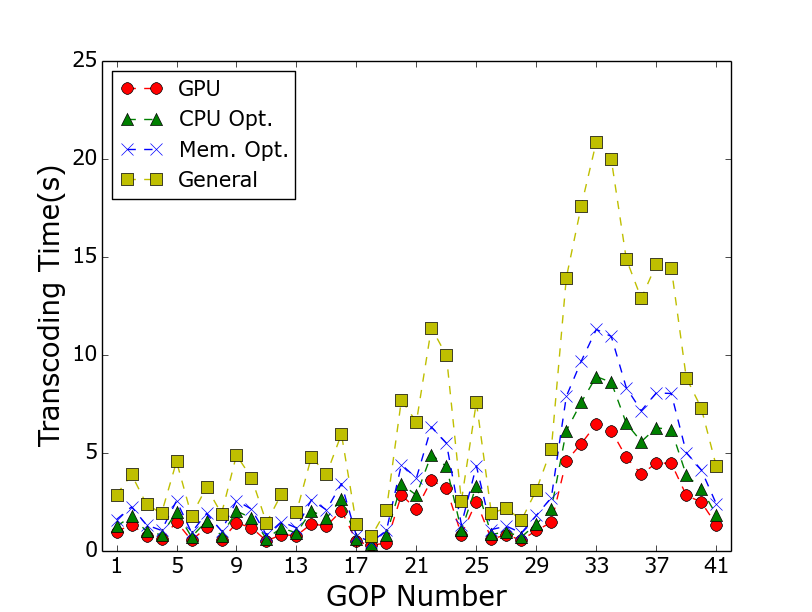}
      \caption{Mixed motion video.}
      \label{fig:mixed}
    \end{subfigure}
    \caption{Transcoding time (in Seconds) of video streams on different cloud VMs for various video content types. (a), (b), and (c) demonstrate the transcoding time obtained from different VM types when applied on slow motion, fast motion and mixed motion video content types, respectively.}
    \label{fig:video_type}
\end{figure*}
As we observed in the previous section, the transcoding time of a GOP can vary significantly on different VM types. For instance, in Figure~\ref{dif_instance}, transcoding time difference between \texttt{GPU} and \texttt{CPU Opt.} VM types for GOP 8 is $\simeq$ 7 seconds while the difference for GOP 13 is less than a half second. What is this performance difference attributed to? Answering this question enables us to allocate the appropriate VM types depending on the GOP type, hence,  reducing the transcoding time and its incurred cost.

Our investigation revealed that the reason for the transcoding time variations is the content type of the GOPs. 
To further investigate the impact of video content type on the transcoding performance, we performed codec transcoding on each video content type on different VM types. Results of the investigation are reported in Figure~\ref{fig:video_type}.

Figure~\ref{fig:slow} shows that the transcoding times of the slow motion videos are distinct from each other across different VM types. In particular, \texttt{GPU} and \texttt{General} VM types, respectively, provide the best and worst performance for this type of video content. 

In contrast, Figure~\ref{fig:fast} shows that the outperformance of \texttt{GPU} VM is not statistically and practically significant when transcoding fast motion videos. Although \texttt{GPU} still provides a slightly faster transcoding time than other VM types, the difference is negligible. For some GOPs (\eg 4, 5, 13, 16, and 31) the transcoding time on GPU is almost the same as other VM types. 

To confirm this finding, we performed the transcoding operation on a mixed motion video and the result is depicted in Figure~\ref{fig:mixed}. As we can see in this sub-figure, \texttt{GPU} outperforms others VMs significantly for some GOPs (\eg. GOP 30 to 37) while provides almost same transcoding time for other GOPs. We noticed that the difference in transcoding time is remarkable for GOPs of the video that contains slow motion content and it is negligible for fast motion GOPs. 

The reason for the performance variations on different video content types is that, in fast motion videos, due to the high frequency of changing scenes, the number of frames in a GOP and, therefore, the GOP size is small. In contrast, slow motion GOPs include more frames and they are larger in size. When we transcode a large number of small size GOPs (\ie the case for fast motion videos) there is little computation to be performed for each GOP and the performance of the VM is dominated by the overhead of switching between different GOPs. On the contrary, when in transcoding slow motion videos we deal with few numbers of GOPs that are large in size (\ie they are compute intensive). Transcoding such videos can take advantage of compute-heavy (\eg \texttt{GPU}) VMs.

In the next section, we will further investigate the impact of GOP size and number of frames in a GOP for video transcoding.

\subsection{Analyzing the Impact of GOP Size and Number of Frames on Transcoding Time}\label{sub:discuss}

In Section~\ref{sub:vt}, we concluded that the transcoding time of GOPs varies significantly on different VMs depending on the video content types. However, automatic categorization of GOPs based on their video content type is a difficult task. We need an intuitive factor to categorize GOPs on different VM types. In this section, we investigate further the factors that influence GOP transcoding time on different VM types.

As we noticed in Section~\ref{sub:vt}, a GOP with slow motion content type benefits more from a computationally powerful VM. Such a GOP has a large size and includes many frames. Therefore, we need to analyze the impact of GOP \emph{size} and \emph{number of frames} on the transcoding time of each GOP on different VM types. 

\begin{figure}[htb] 
    \centering
    \includegraphics[width=3.5in]{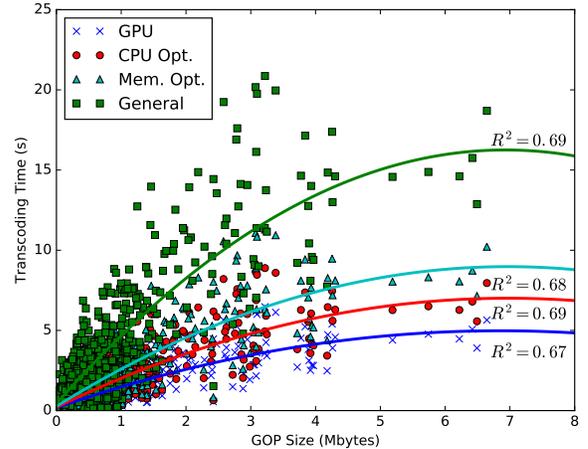}
    \caption{Second degree regression to study the influence of GOP size on the transcoding time.}
    \label{fig:tgs}
\end{figure}

\begin{figure}[htb] 
    \centering
    \includegraphics[width=3.5in]{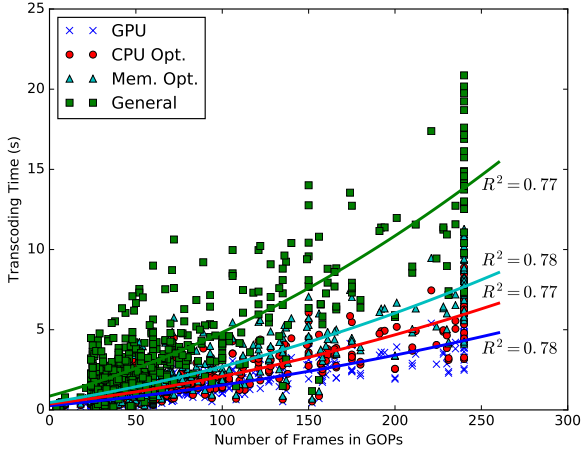}
    \caption{Second degree regression to study the influence of number of frames in a GOP on the transcoding time.}
    \label{fig:tfn}
\end{figure}

We use a regression analysis to study the impact of GOP size and number of frames on the GOP transcoding time. We consider the transcoding time of GOPs in all benchmark videos of the benchmark that is a mixture of slow, fast, and mixed motion video contents. Due to the large amount of data, the second-degree regression is used for the analysis. The horizontal axis shows The GOP size (in MB) and number of frames for all GOPs in Figures~\ref{fig:tgs} and~\ref{fig:tfn}, respectively. The vertical axes show the transcoding times of GOPs (in seconds).

In both figures, we observe that, regardless of the VM type, transcoding times increase by increasing both the GOP size and the GOP number of frames. As we can, the coefficient of determination ($R^2$) for the regression analyses. As we can see in this table, both GOP size and number of frames show a high confidence of relationship to transcoding time, while the number of frames in a GOP shows a higher $R^2$ value for all VM types. Therefore, the number of frames provides a stronger regression with transcoding time. 

In Figure~\ref{fig:tfn}, we also observe that when the number of frames in a GOP is small, the performance of \texttt{GPU} is very close to other VM types whereas for a larger number of frames, the performance gap between GPU and others VM types rises. This implies that GOPs with few numbers of frames are better to be assigned to cost-efficient VM types whereas GOPs with a large number of frames can benefit from computationally powerful VM types. 



\section{Performance Cost Trade-Off of Transcoding on Heterogeneous VMs}\label{sec:md}
\subsection{Overview}
VM types offered by cloud providers are heterogeneous both in terms of performance and cost~\cite{dillon2010cloud}. Hence, allocating VMs that are cost- and performance-efficient for transcoding tasks is challenging. 

As we discussed in Section~\ref{sub:it}, computationally-powerful VMs do not always provide the best performance for transcoding tasks. This is particularly important when we consider the significant cost difference between the VM types. We also discussed that the transcoding time has a correlation with the GOP size and number frames in GOPs. In particular, when the GOP size or number of frame is small, the performance difference of heterogeneous VMs is negligible.
Alternatively, the performance difference of using heterogeneous VMs to transcode large size GOPs is significant. Thus, it may be worthwhile to allocate a powerful and costly VM to transcode such GOPs. 

To cope with the appropriate VM type allocation challenge, we require a construct to identify the appropriateness of various VM types for different GOPs. Such a construct can be helpful in allocation and mapping (\ie scheduling) of GOPs to the appropriate VMs for transcoding. In this section, we present a construct termed GOP \emph{Suitability Matrix} that maintains the suitability value of each VM type for each GOP task in a video stream. Such a matrix can be used by video stream providers to allocate VMs that offer the best performance and cost trade-off for video transcoding.

\subsection{Modeling Performance Cost Trade-Off of Transcoding Tasks on Heterogeneous VMs} 
Recall from Table~\ref{tbl:cost-table} that \texttt{GPU} VM type, in general, provides the best performance while having the highest cost. Also, \texttt{General} VM type provides the lowest transcoding performance and is the least expensive one when compared to other VMs. 

We define \emph{performance gap}, denoted $\Delta_i$, as the performance difference VM type $i$ and \texttt{GPU}. For a given GOP, a large value of $\Delta_i$ indicates that VM type $i$ remarkably performs worse than \texttt{GPU}, hence, \texttt{GPU} should be assigned a higher suitability value than VM $i$. 

Determination of the trade-off between performance and cost of utilizing heterogeneous VMs, in the first place, depends on the business policy of the streaming service provider (here, we call it \emph{user}). That is, a user should determine how important is the performance, denoted $p$, and the incurred cost, denoted $c$, for the system. As these parameters complement each other (\ie $p + c = 1$), the user only needs to provide one of these parameters. For instance, a user can provide $p=0.6$ (that implies $c=0.4$) to indicate a higher performance preference.

\begin{figure}[htb] 
    \centering
    \includegraphics[width=3.5in]{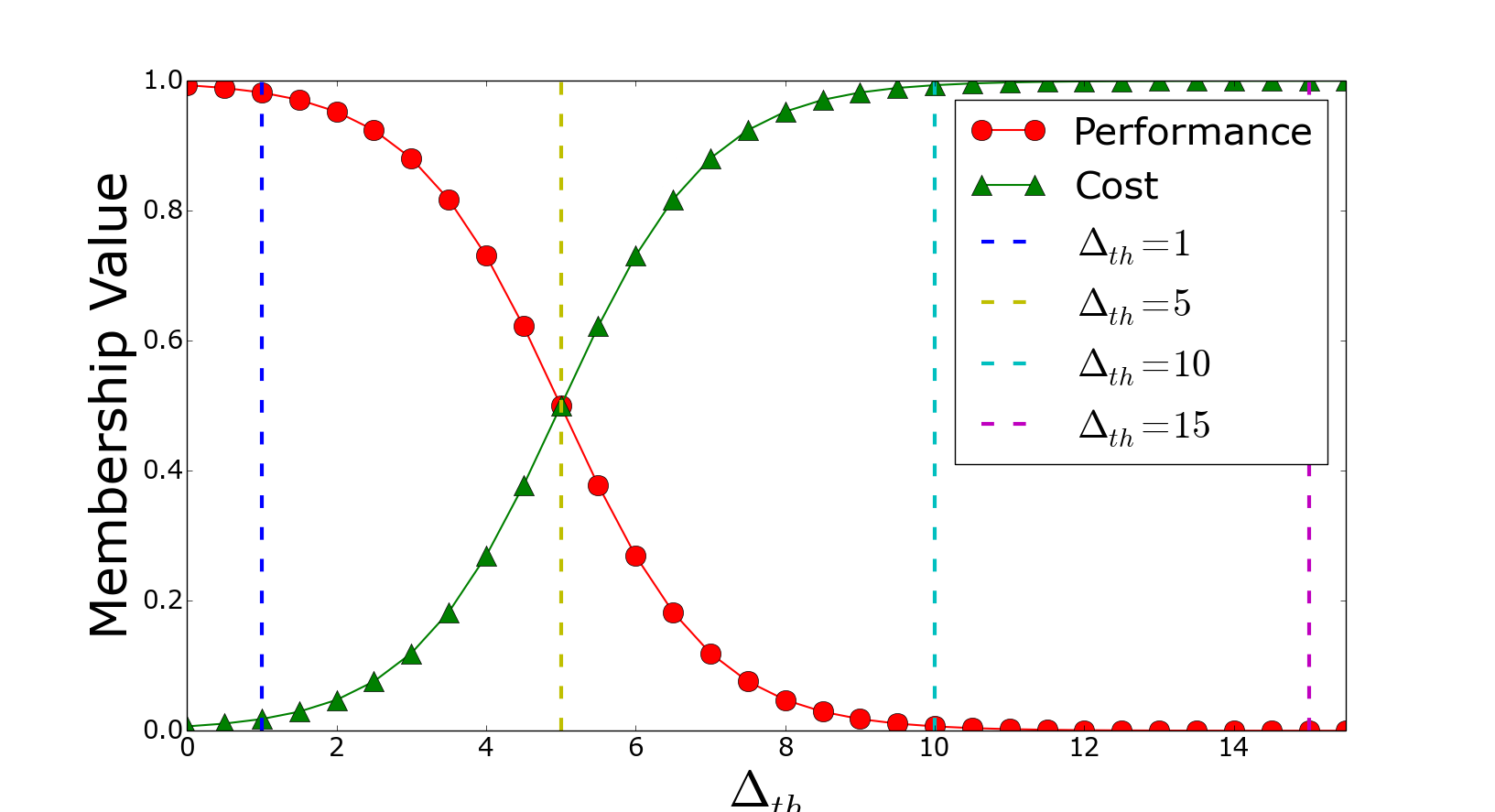}
    \caption{Membership functions for performance and cost preferences. The user-provided values for the cost or performance are considered as the membership value (vertical axis). Then, the corresponding value on the horizontal axis is considered as $\Delta_{th}$}
    \label{fig:fuzz}
\end{figure}

We define \emph{performance threshold gap}, denoted $\Delta_{th}$, as the threshold of the performance gap between \texttt{GPU} and other VM types. The value of $\Delta_{th}$ is determined based on the user preference of $p$ and $c$. As user cost and performance preferences are not crisp values, we can model them based on fuzzy membership functions~\cite{puri1983differentials}. As shown in Figure~\ref{fig:fuzz}, we define two membership functions for the cost and performance preferences. According to this figure, the membership value of one preference (\eg performance) decreases when the other preference (\eg cost) increases. 

Value of the user's performance (or cost) preference is considered as the membership value of the fuzzy membership function (vertical axis in Figure~\ref{fig:fuzz}) and is used to obtain the performance threshold gap (horizontal axis in Figure~\ref{fig:fuzz}). More specifically, by using the performance preference ($p$), we can obtain $\Delta_{th}$ based on Equation~(\ref{eq:fuzzy_p})\footnote{Similarly, the value of $\Delta_{th}$ can be obtained from the cost preference value: $\Delta_{th} = \frac{\ln{\frac{c}{1-c}}}{\alpha} - \beta$}.

\renewcommand{\arraystretch}{1.5}
\begin{table*}[htb]
    \begin{subtable}[h]{0.48\textwidth}
        \centering
        \resizebox{\columnwidth}{!}{
          \begin{tabular}{ c | c | c | c | c }
           VM Type & \texttt{General} & \texttt{CPU Opt.} & \texttt{Mem. Opt.} & \texttt{GPU} \\ \hline
           $GOP_1$ & {\bf0.00} & 1.00 & 0.78 & {\bf0.98}\\
           $GOP_2$ & {\bf0.00} & 1.00 & 0.68 & {\bf0.26}\\
           $GOP_3$ & {\bf0.00} & 1.00 & 0.67 & {\bf0.30}\\ 
           $GOP_4$ & {\bf0.00} & 1.00 & 0.61 & {\bf0.01}\\ 
           $GOP_5$ & {\bf0.00} & 1.00 & 0.71 & {\bf0.60}\\ 
           $GOP_6$ & {\bf0.00} & 1.00 & 0.80 & {\bf0.89}\\ 
           $GOP_7$ & {\bf0.00} & 0.91 & 0.74 & {\bf1.00}\\ 
           $GOP_8$ & {\bf0.00} & 0.88 & 0.72 & {\bf1.00}\\ 
           $GOP_9$ & {\bf0.00} & 0.87 & 0.72 & {\bf1.00}\\ 
        $GOP_{10}$ & {\bf0.00} & 0.86 & 0.71 & {\bf1.00} \\ 
           \vdots & \vdots & \vdots & \vdots & \vdots \\ 
          \end{tabular}
        }
        \caption{Suitability Matrix, when $p$ = 98\%, $c$ = 2\%, and $\Delta_{th}$ = 1}
        \label{tbl:suit-table-1}
    \end{subtable}
    \hfill
    \begin{subtable}[h]{0.48\textwidth}
        \centering
        \resizebox{\columnwidth}{!}{
          \begin{tabular}{ c | c | c | c | c }
           VM Type & \texttt{General} & \texttt{CPU Opt.} & \texttt{Mem. Opt.} & \texttt{GPU} \\ \hline
           $GOP_1$ & 0.00 & 1.00 & 0.63 & 0.03\\
           $GOP_2$ & 0.69 & 1.00 & 0.78 & 0.00\\
           $GOP_3$ & 0.67 & 1.00 & 0.78 & 0.00\\ 
           $GOP_4$ & 0.75 & 1.00 & 0.78 & 0.00\\ 
           $GOP_5$ & 0.57 & 1.00 & 0.74 & 0.00\\ 
           $GOP_6$ & 0.26 & 1.00 & 0.71 & 0.00\\ 
           $GOP_7$ & 0.00 & 1.00 & 0.72 & 0.54\\ 
           $GOP_8$ & 0.00 & 1.00 & 0.75 & 0.70\\ 
           $GOP_9$ & 0.00 & 1.00 & 0.77 & 0.77\\ 
        $GOP_{10}$ & 0.00 & 1.00 & 0.77 & 0.77\\ 
           \vdots & \vdots & \vdots & \vdots & \vdots \\ 
          \end{tabular}
        }
        \caption{Suitability Matrix, when $p$ = 50\%, $c$ = 50\%, and $\Delta_{th}$ = 5}
        \label{tbl:suit-table-5}
    \end{subtable}
    \begin{subtable}[h]{0.48\textwidth}
        \vspace*{0.8cm}
        \centering
        \resizebox{\columnwidth}{!}{
          \begin{tabular}{ c | c | c | c | c }
           VM Type & \texttt{General} & \texttt{CPU Opt.} & \texttt{Mem. Opt.} & \texttt{GPU} \\ \hline
           $GOP_1$ & 0.48 & 1.00 & 0.73 & 0.00 \\
           $GOP_2$ & 0.79 & 1.00 & 0.80 & 0.00 \\
           $GOP_3$ & 0.79 & 1.00 & 0.80 & 0.00 \\ 
           $GOP_4$ & 0.83 & 1.00 & 0.80 & 0.00 \\ 
           $GOP_5$ & 0.74 & 1.00 & 0.78 & 0.00 \\ 
           $GOP_6$ & 0.59 & 1.00 & 0.77 & 0.00 \\ 
           $GOP_7$ & 0.06 & 1.00 & 0.65 & 0.00 \\ 
           $GOP_8$ & 0.00 & 1.00 & 0.68 & 0.20 \\ 
           $GOP_9$ & 0.00 & 1.00 & 0.71 & 0.35 \\ 
        $GOP_{10}$ & 0.00 & 1.00 & 0.70 & 0.32 \\  
           \vdots & \vdots & \vdots & \vdots & \vdots \\ 
          \end{tabular}
        }
        \caption{Suitability Matrix, when $p$ = 1\%, $c$ = 99\%, and $\Delta_{th}$ = 10}
        \label{tbl:suit-table-10}
    \end{subtable}
    \hfill
    \begin{subtable}[h]{0.48\textwidth}
        \vspace*{0.8cm}
        \centering
        \resizebox{\columnwidth}{!}{
          \begin{tabular}{ c | c | c | c | c }
           VM Type & \texttt{General} & \texttt{CPU Opt.} & \texttt{Mem. Opt.} & \texttt{GPU} \\ \hline
           $GOP_1$ & {\bf0.63} & 1.00 & 0.76 & {\bf0.00} \\
           $GOP_2$ & {\bf0.82} & 1.00 & 0.81 & {\bf0.00} \\
           $GOP_3$ & {\bf0.83} & 1.00 & 0.81 & {\bf0.00} \\ 
           $GOP_4$ & {\bf0.85} & 1.00 & 0.81 & {\bf0.00} \\ 
           $GOP_5$ & {\bf0.79} & 1.00 & 0.80 & {\bf0.00} \\ 
           $GOP_6$ & {\bf0.69} & 1.00 & 0.79 & {\bf0.00} \\ 
           $GOP_7$ & {\bf0.37} & 1.00 & 0.72 & {\bf0.00} \\ 
           $GOP_8$ & {\bf0.19} & 1.00 & 0.69 & {\bf0.00} \\ 
           $GOP_9$ & {\bf0.04} & 1.00 & 0.66 & {\bf0.00} \\ 
        $GOP_{10}$ & {\bf0.08} & 1.00 & 0.67 & {\bf0.00} \\  
           \vdots & \vdots & \vdots & \vdots & \vdots \\ 
          \end{tabular}
        }
        \caption{Suitability Matrix, when $p$ = 0.01\%, $c$ = 99.99\%, and $\Delta_{th}$ = 15}
        \label{tbl:suit-table-15}
    \end{subtable}
    \vspace*{0.2cm}
    \caption{Suitability Matrices for different values of performance and cost preferences. Tables (a) to (d),  show that as the performance preference $p$ decrease (and cost-preference $c$ increases), the value of $\Delta_{th}$ grows. Accordingly, the maximum Suitability value changes from \texttt{GPU} (performance-oriented VM) in Table~\ref{tbl:suit-table-1} to \texttt{General} type (cost-oriented VM) in Table~\ref{tbl:suit-table-15}.}
    \label{tbl:suit-matrix}
\end{table*}

\begin{equation}\label{eq:fuzzy_p}
      \Delta_{th} = \frac{\ln{\frac{1-p}{p}}}{\alpha} + \beta
\end{equation}

where $\alpha$ is the inflection point in the membership function and $\beta$ is the slope at $\alpha$. In Figure~\ref{fig:fuzz}, we experimentally obtained the values of $\alpha$ and $\beta$ equal to 1 and 5, respectively. 

Based on the value of $\Delta_{th}$, we can determine the trade-off between performance and cost for transcoding a given GOP. For that purpose, we define \emph{weight} of the VM type $i$, denoted $W_i$, to transcode a given GOP based on Equation~\ref{eq:suitability} that encompasses both the performance and cost factors. 

The first part, in Equation~\ref{eq:suitability}, considers the performance factor and calculates the  difference of performance gap from $\Delta_{th}$. Performance gaps greater than the threshold ($\Delta_{th}$) cause a low (negative) weight value which implies higher Suitability for performance-oriented VM types. In this part, the denominator determines the sum of performance gaps for all $N$ VM types. The second part, in Equation~\ref{eq:suitability}, considers the cost factor. This part functions based on the cost of transcoding a given GOP on VM type $i$, denoted $\varphi_i$. The cost of transcoding a GOP on VM $i$ is obtained from the transcoding time of the GOP on VM $i$ and the hourly cost of VM type $i$ in the cloud. This part of the equation favors VM types that incur a lower cost for transcoding a given GOP.

\begin{figure*}[htbp]
    \centering
    \begin{subfigure}[b]{0.32\textwidth}
      \centering
      \includegraphics[width=\textwidth]{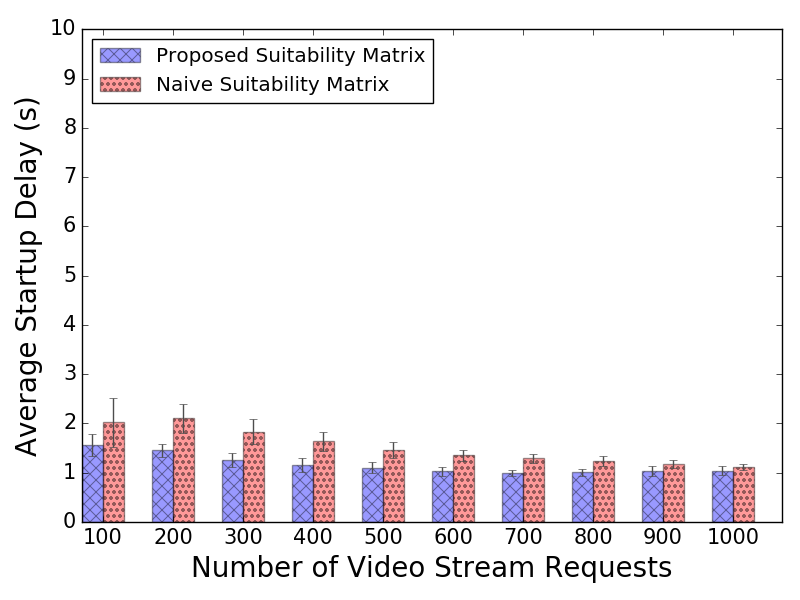}
      \caption{Comparison of startup delay}
      \label{fig:stt_comp}
    \end{subfigure}
    \begin{subfigure}[b]{0.32\textwidth}
      \centering
      \includegraphics[width=\textwidth]{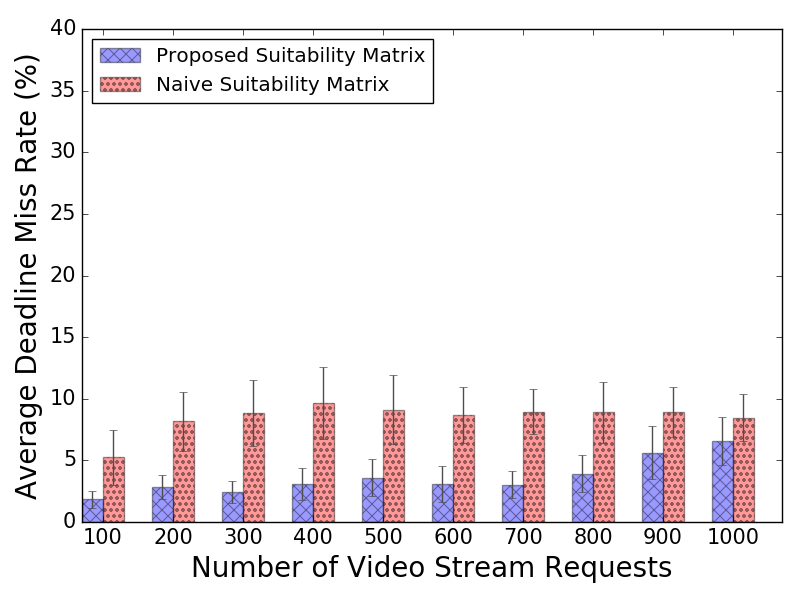}
      \caption{Comparison of deadline miss rate}
      \label{fig:dmr_comp}
    \end{subfigure}
    \begin{subfigure}[b]{0.32\textwidth}
      \centering
      \includegraphics[width=\textwidth]{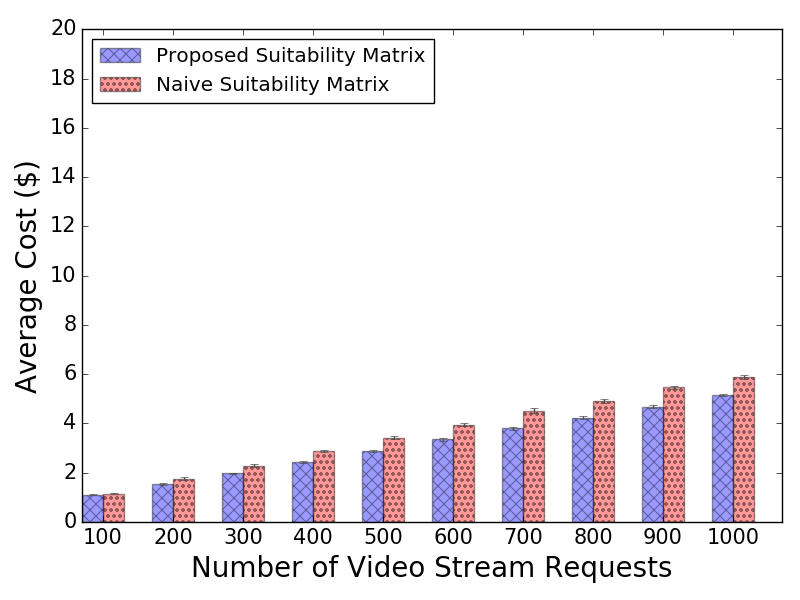}
      \caption{Comparison of cost}
      \label{fig:cost_comp}
    \end{subfigure}
    \caption{Performance and cost comparison when our proposed suitability matrix is used against the a na\"ive suitability matrix. Horizontal axes in all subfigures show the number of streaming tasks.}
    \label{fig:tpds_vs_analysis}
\end{figure*}

\begin{figure*}[htbp]
    \centering
    \begin{subfigure}[b]{0.32\textwidth}
      \centering
      \includegraphics[width=\textwidth]{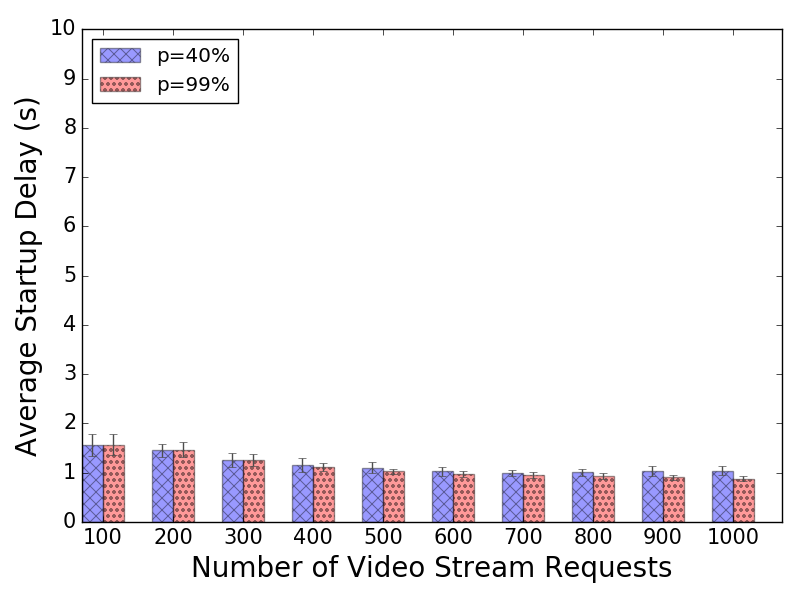}
      \caption{Comparison of startup delay}
      \label{fig:stt_comp_analysis}
    \end{subfigure}
    \begin{subfigure}[b]{0.32\textwidth}
      \centering
      \includegraphics[width=\textwidth]{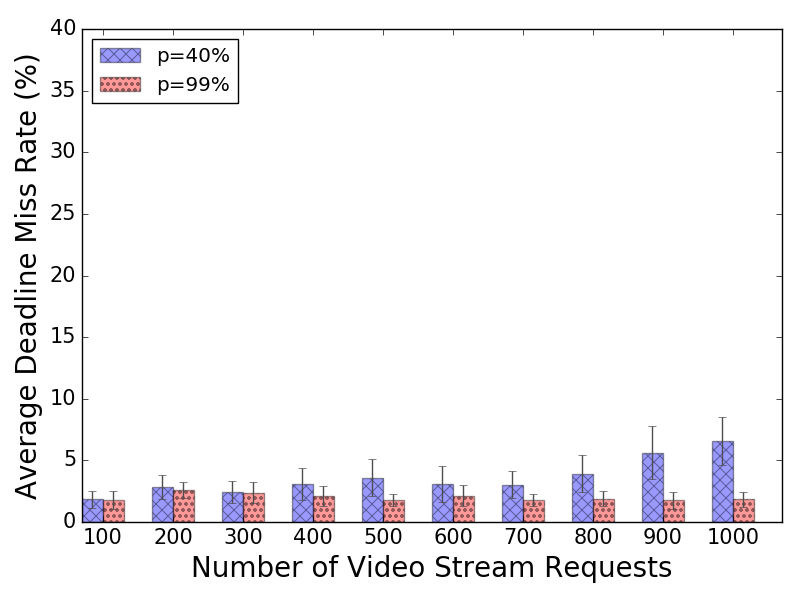}
      \caption{Comparison of deadline miss rate}
      \label{fig:dmr_comp_analysis}
    \end{subfigure}
    \begin{subfigure}[b]{0.32\textwidth}
      \centering
      \includegraphics[width=\textwidth]{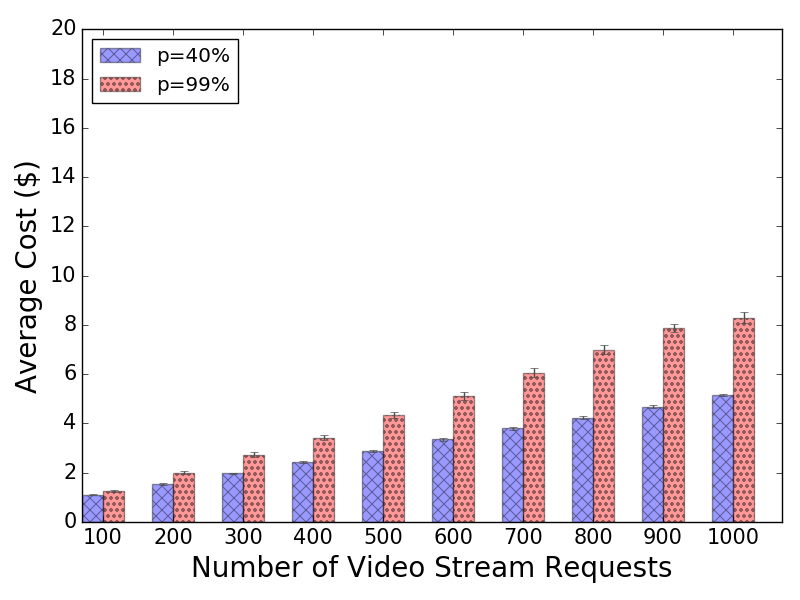}
      \caption{Comparison of incurred cost}
      \label{fig:cost_comp_analysis}
    \end{subfigure}
    \caption{Performance and cost comparison of our proposed suitability matrix with different performance rates, $p$=40\% versus $p$=99\%. Horizontal axes in all subfigures show the number of streaming tasks.}
    \label{fig:comp_analysis}
\end{figure*}

\begin{equation}\label{eq:suitability}
    W_i= \frac{\Delta_{th}-\Delta_i}{\sum\limits_{n=1}^{N} \Delta_n} \cdot (1-\frac{\varphi_i}{\sum\limits_{n=1}^{N} \varphi_n})
\end{equation}

To normalize the value of $W_i$ and determine the final suitability values, denoted $S_i$, between [0, 1] we use Equation~(\ref{eq:nomalization}) as follows: 

\begin{equation}\label{eq:nomalization}
    S_i= \frac{W_i-\max\limits_i(W_i)}{\max\limits_i(W_i) - \min\limits_i(W_i)}
\end{equation}

where $max_{i}$ and $min_{i}$ are the largest and smallest values among $W_i$s, respectively. 
In transcoding a video stream, each GOP has  different Suitability values on different VM types. These suitability values construct a \emph{Suitability Matrix} for each video stream. 

\subsection{Case Study of the Trade-Off Model}
To have a better understanding of the Suitability Matrix construct, we compare four suitability matrices with different performance and cost preference values provided by the user. 

Table~\ref{tbl:suit-table-1} shows the Suitability Matrix for a given video when the user has a performance-oriented preference---$p=0.98$. As we can see, in this case, the Suitability value of \texttt{GPU} and \texttt{CPU Opt.} VMs is higher than the other VM types. We observe that the Suitability values for (\texttt{General}) VM are mostly 0.

When user's performance preference drops to 0.5 (and cost raises to 0.5), as demonstrated in Table~\ref{tbl:suit-table-5}, the Suitability value of \texttt{GPU} decreases while the \texttt{General} VM gets higher Suitability values. By further decreasing the user performance preference and increasing the cost preference, the Suitability value of the \texttt{GPU} drops to almost 0 while the Suitability values of cost-efficient VMs (\texttt{General}) are increased (see Tables~\ref{tbl:suit-table-10} and~\ref{tbl:suit-table-15}). It is noteworthy that \texttt{CPU Opt.} VM type mostly maintains a high Suitability value regardless of $\Delta_{th}$ value. This is because the \texttt{CPU Opt.} VMs has a high performance and its cost is relatively low. 

\subsection{Performance Evaluation}
In the experiments of this section, we used CloudSim~\cite{cloudsim}, a discrete event simulator, to model our system and evaluate performance of the scheduling methods and VM provisioning policies. We modeled the system based on the characteristics and cost of VM types in Amazon EC2. We measured the startup delay, deadline miss rate of video streams, and the incurred cost of using cloud VMs to process different number of streaming tasks (from 100 or 1000) arriving during the same time period\footnote{Details of the generated workload can be downloaded from https://goo.gl/TE5iJ5}. For the sake of accuracy, each experiment has been conducted 30 times and the mean and 95\% confidence interval of the results are reported. For this experiment, we consider the performance ratio $p$=40\% (and cost ratio $c$=60\%).

To demonstrate the efficacy of our proposed trade-off model, in the first experiment, we compare the performance and the incurred cost when the scheduling method uses the proposed suitability matrix against a na\"ive suitability matrix that has been proposed in~\cite{li2017cost}. 

The na\"ive method operates simply based on a trade-off between the performance ($T_i$) and the cost ($C_i$) for a given GOP on VM type $i$, as shown in Equation (\ref{eq:weight}), while it does not consider the performance tolerance that user can decide like our proposed approach.
\begin{equation}\label{eq:weight}
      S_i =  k \cdot T_i + (1-k) \cdot C_i
\end{equation}

As we can see in Figure~\ref{fig:tpds_vs_analysis}, the resource allocation system that uses our proposed suitability matrix leads to a lower startup delay and a lower deadline miss rate at even a lower cost. The reason is our proposed method can more accurately assign GOP types based on user's preference.

To further investigate the impact of SSP's preference on the performance (and cost) when our proposed suitability matrix is deployed, in the second experiment, we compared the performance and the incurred cost with two performance ratios, namely $p$=40\% and $p$=99\%. Figure~\ref{fig:comp_analysis} expresses that for the higher value of performance ratio, both the startup delay and deadline miss rate is improved. The improvement is more remarkable when there are more tasks in the system. In addition, we can see that the incurred cost also significantly increases for a higher performance ratio. The experiment testifies that the  performance and incurred cost resulted from deploying the proposed suitability matrix conforms with the discretion of the streaming service provider.

\section{Related Work}\label{sec:rw}

Several studies explored the performance analysis of heterogeneous cloud services~\cite{rw_21, bg_15, bg_16}. Iosup~\etal~\cite{rw_21} and Jackson~\etal~\cite{bg_16} studied application-oriented performance analysis using heterogeneous cloud services. The results show that although cloud services have their own drawbacks in terms of communication and processing delays, utilizing cloud services is a viable solution for processing workloads that need resources instantly and temporarily. Lee \etal~\cite{bg_15} investigate the task-machine affinity in heterogeneous clusters. They propose a shared metric in the heterogeneous cluster to provide a scheduling method that considers fairness. However, there is no study in the literature that focuses on analyzing video transcoding tasks on heterogeneous VM types in clouds.

\emph{Expected Time to Compute} (ETC)~\cite{rw_17, intro_13, intro_14} and \emph{Estimated Computation Speed} (ECS)~\cite{intro_12, intro_15} matrices are commonly used to explain the affinity of different tasks types on heterogeneous machines. These matrices are utilized for more efficient task scheduling and VM allocation. However, the definition of both ETC and ECS only considers execution time as the performance metric and ignores the cost heterogeneity across different VM types. Our proposed \emph{Suitability Matrix} extends the idea of ETC matrices by including both performance and cost metrics.

Video transcoding is a computationally expensive and time-consuming operation. Techniques, architectures, and the challenges of video transcoding were investigated by Ahmad \etal~\cite{intro_6} and Vetro \etal~\cite{intro_7}. With the rise of cloud computing, Streaming Service Providers (SSPs) realize a more cost-efficient way to transcode videos by utilizing cloud services. 

\begin{figure}[htb] 
    \centering
    \includegraphics[width=3.5in]{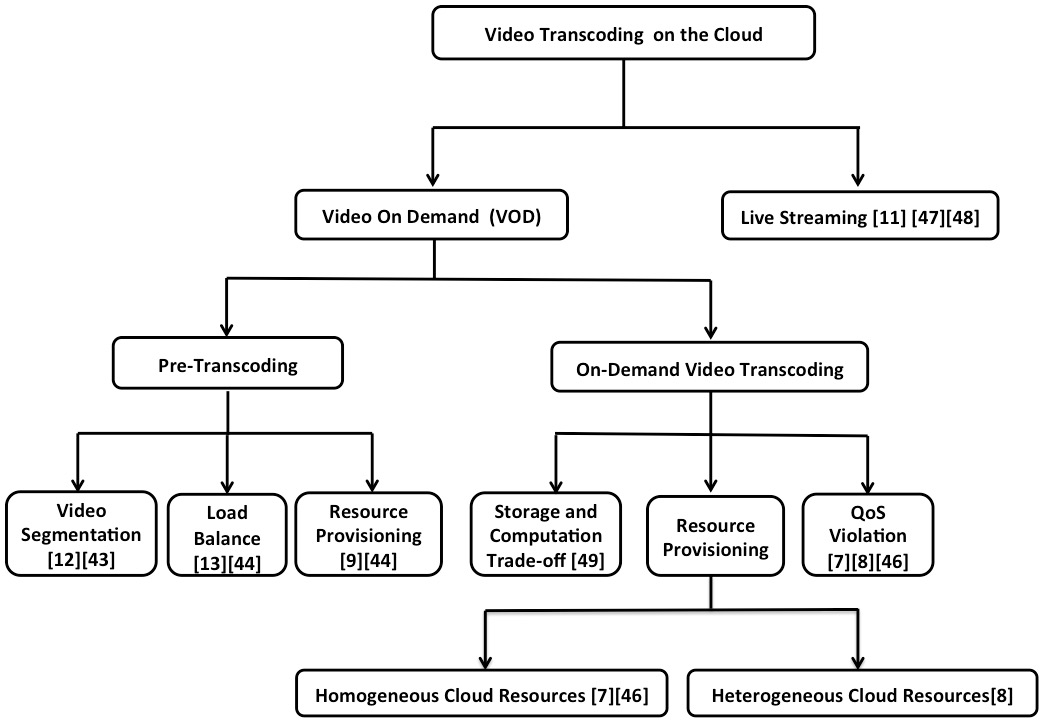}
    \caption{A taxonomy of researches undertaken on video transcoding using cloud services.}
    \label{fig:rw}
\end{figure}

A taxonomy of the studies undertaken on cloud-based video transcoding is illustrated in Figure~\ref{fig:rw}. Challenges of cloud-based transcoding for VOD was studied in~\cite{rw_10, rw_11}. Studies have been concentrated on both pre-transcoding~\cite{rw_12, bg_2, rw_3, rw_10, rw_11}, on-demand transcoding~\cite{pre_3, pre_4, li2012cloud} and live streaming~\cite{pre_5, rw _22, thang2012adaptive}.

For pre-transcoding, the research focus is mainly on video segmentation~\cite{rw_12, bg_2}, load balance~\cite{rw_3, rw_11}, and resource provisioning~\cite{rw_3, rw_10}, while the quality of service (QoS) is not a concern because different versions of the same video will be ready before releasing to viewers. However, transcoding the whole repository videos into multiple versions and storing all the versions causes massive storage cost for SSPs.

To reduce the storage cost while remaining QoS, on-demand video transcoding has been proposed in~\cite{pre_3, li2017cost,li2012cloud}. Li \etal~\cite{pre_3, li2017cost} propose the CVSS architecture to efficiently transcode video in an on-demand manner on homogeneous and heterogeneous cloud VMs, respectively. With proper scheduling and resource provisioning policy, CVSS provides low startup delay and playback jitter. Li \etal~\cite{li2012cloud} present a Cloud Transcoder which utilizes an intermediate cloud platform to bridge the format/resolution gap for mobile devices in real-time. It only requires the user to upload a video request with specified transcoding parameters rather than the video content. Cloud Transcoder transcodes downloads and transcodes the original video on the user's demand and deliver the transcoded version the user.

Jokhio \etal~\cite{rw_13} presents a computation and storage trade-off strategy for cost-efficient video transcoding in the cloud. The trade-off is based on the computation cost versus the storage cost of the video streams. They determine how long a video should be stored or how frequently it should be re-transcoded from a given source video. Zhao~\etal~\cite{rw_18} take the popularity, computation cost, and storage cost of each version of a video stream into account to determine versions of a video stream that should be stored or transcoded. Kathpal \etal \cite{atish12} developed cost metrics that enable comparing storage versus compute costs and determine when an on-demand transcoding can be cost-effective. They also analyze how such a solution can be deployed in a storage system based on the access pattern information or online algorithms when such access patterns are not available.
 
The idea of cloud-based video transcoding has also has been applied to live video streaming~\cite{pre_5, rw _22}. Timmerer\etal~\cite{timmerer2015live} present a live transcoding and streaming-as-a-service architecture utilizing cloud infrastructure taking the live video streams as input and output multiple stream versions according to the MPEG-DASH~\cite{thang2012adaptive} standard. Lai \etal~\cite{lai2013cloud} design a cloud-assisted real-time transcoding mechanism based on the HLS protocol~\cite{stockhammer2011dynamic}, they implement the bandwidth recoder, segment transrater, and segment redirector on the server. They provide an instant analysis of the online quality between client and server without changing the HLS server architecture and the optimum media quality.

With the trend of video transcoding using cloud services, a better understanding the performance of different video transcoding operation on heterogeneous VMs is necessary. Transcoding time estimation plays an important role in both efficient scheduling and resource provisioning. Deneke \etal~\cite{deneke2014video} utilize machine learning methods based on the video characteristics (\eg resolution, frame rate, and bit rate) to predict the transcoding time. Seo \etal~\cite{seo2010load} focus on the transcoding process details to estimate transcoding time, such as discrete cosine transform (DCT), inverse DCT (iDCT), quantization (Q), inverse Q (iQ), motion estimation/motion compensation (ME/MC), variable length coding (VLC), variable length decoding (VLD). While both~\cite{deneke2014video, seo2010load} do not consider the diversity of heterogeneous environment of cloud services. Our work provides a deep performance analysis and transcoding time estimation for different transcoding operations on heterogeneous VM types, which is beneficial for cost and performance efficient video transcoding scheduling and resource provisioning using cloud services.

\section{Summary and Discussion}\label{sec:conclusion}
With the emergence of on-demand video transcoding on the cloud, it is crucial to study the video transcoding tasks and influential factors on their execution times. In addition, it is necessary to come up with a trade-off between performance and cost of using cloud services. The trade-off becomes further complicated when we consider the heterogeneity of computational services (VMs) offered by cloud providers. To understand the affinity of different transcoding tasks and heterogeneous VM types we provided a detailed study and analysis of different transcoding operations on heterogeneous VMs. In summary, the main findings of our research are as follows:
\begin{enumerate}
\item The execution times of different transcoding operations follow a pattern:  video codec and adjusting frame rate transcoding require more computation time than bit rate and spatial resolution transcoding.

\item Although \texttt{GPU} VM type mostly provides a faster execution time than other VM types, in some cases the execution time difference is negligible. In particular, we observed that when transcoding tasks are categorized based on transcoding type, up to 63\% of bit rate transcoding tasks can be executed on VM types other than \texttt{GPU} with nearly the same transcoding time (see Table~\ref{tbl:perf-table-2}) while incurring a significantly lower cost. 

\item We learned that the execution time of the transcoding operation on heterogeneous VMs has a correlation with the video content type. GOPs that contain slow motion video content are larger in size and include more frames in compare to GOPs of fast motion videos. Thus, GOPs with slow motion video content can benefit from computationally powerful VMs whereas fast motion ones can be executed on less powerful and more cost efficient VMs with a similar performance. 

\item Cloud VMs exhibit inconsistent heterogeneity behavior in executing video transcoding tasks. However, the inconsistent behavior is more related to video content type rather than the type of transcoding operation. As such, video transcoding tasks (GOPs) are suggested to be categorized based on their content type to gain more from heterogeneous VMs offered by cloud providers.  

\item As identifying GOPs' content types prior to execution is difficult, we can use the number of frames (or frame size) in the GOP as an intuitive factor that indicates the content type of transcoding tasks.

\item By considering both the performance and cost heterogeneity of different VM types, we provided a model that, identifies the degree of suitability of each VM type for a given GOP. The provided model operates based on the SSP performance and cost preference. Suitability matrices can supply resource allocation and scheduling methods with accurate performance and cost trade-offs to utilize appropriate VMs for video transcoding.

\item Evaluations show that in comparison to na\"ive method in~\cite{li2017cost}, our suitability matrix provides a lower startup delay and a lower deadline miss rate at a lower cost. 
\end{enumerate}


\bibliographystyle{IEEEtran}
\bibliography{paper}

\end{document}